\documentstyle[12pt,epsfig]{article}
\begin{document}
\sloppy
\begin{titlepage}
\samepage{
\setcounter{page}{0}
\vspace{-3.in}
\rightline{DFTT 36/97}
\rightline{UB-ECM-PF 97/08}
\rightline{hep-ph/9706390}
\vspace{.1in}
\begin{center}
{\bf PROGRESS IN PERTURBATIVE QCD\\}
\vspace{.3in}
{Stefano Forte\\}
\vspace{.15in}
{\it INFN, Sezione di Torino,\\ Via P.~Giuria 1, I-10125 Torino, Italy\\}
\vspace{.1in}
{\rm and}\\
\vspace{.1in}
{\it Departament ECM, Universitat de
Barcelona,\footnote{IBERDROLA visiting professor}\\
Diagonal 647, E-08028 Barcelona, Spain}
\end{center}
\vspace{.3in}

\begin{abstract}

We briefly summarize some recent theoretical developments in
 perturbative QCD, emphasizing new ideas which
have led to  widening  the domain of applicability
of perturbation theory. In particular, it is now possible to
calculate efficiently processes with many 
partons;  the high order behavior of perturbation theory can be at
 least partly
 understood by going beyond leading twist accuracy; factorization  
with more than one 
hard scale
(such as in DIS with heavy quarks) can be made consistent with the
renormalization group; and large infrared logs can 
be resummed beyond
the renormalization group. The use of the renormalization
 group  to resum
large longitudinal scales may allow the use of
perturbation theory   even in the absence of a large transverse scale.

\end{abstract}
\vspace{.6in}
\begin{center}
Summary of the theory working group\\ at {\bf DIS 97},
Chicago, April 1997\\
\vspace{.1in}
{\it To be published in the proceedings}
\end{center}
\vfill
%\leftline{CERN-TH/95-184}
\leftline{June 1997}
}
\end{titlepage}
\section*{Perturbative QCD beyond leading order}
It is common knowledge that within the last decade the status of
perturbative QCD has changed from being a part of the standard model
which could be only qualitatively tested  to that
of a firmly established theory which has been tested to high
accuracy. The theory can now
be used to perform the detailed calculations required, for instance,
in order to determine the background to new physics --- as it is the case
for the recent HERA
high $Q^2$ events, where QCD uncertainties can be traced and
quantitatively assessed in detail\cite{kuhlm}.
From the theoretical side, this is due to the fact that a set of
standard perturbative results and techniques is now available in order to
describe within a unified framework
numerous phenomena, ranging from the completely inclusive (such as
scaling violations of structure functions) all the way  to
observables related to
completely exclusive parton production processes (such as the momentum
distribution of large $p_T$ jets).

Broadly speaking, perturbative QCD computations are made possible by
factorizing the cross section
$\sigma(P_1,\dots,P_n)$ for a
process with $n$ (one or two, in physically relevant cases) 
hadrons in the initial state in terms of  parton
distributions $f_{i_k}$ for partons $i$ in the $k$-th hadron, and a
hard partonic cross section $\hat\sigma{i_1,\dots,i_n}$: \cite{sterfac}
\begin{equation}
\sigma(P_1,\dots,P_n)=\sum_{i_1,\dots,i_n}\int dx_1\dots dx_n
f_{i_1}(x_1,\mu^2)\dots f_{i_n}(x_n,\mu^2)\hat\sigma_{i_1,\dots i_n}(x_1 P_1,\dots,x_n P_n;\alpha(\mu^2) ).
\label{efact}
\end{equation}
Standard techniques involve the determination of the hard cross section
from perturbative calculation of Feynman diagrams in
powers of $\alpha(\mu^2)$, factorization of   collinear singularities
in the parton distributions,  and renormalization-group improvement of
the dependence of the latter on the factorization  scale $\mu^2$ by
solving the QCD evolution equations. Setting the otherwise
arbitrary scale $\mu^2$ equal to a physical scale of the process then 
ensures that the perturbative expansion of $\hat \sigma$ is free of
large logarithms.
However, the dependence on the arbitrary  scale 
$\mu$ can only be reduced by performing these computations to 
high enough order. The state of the art in most precision tests of QCD
is based on next-to-leading order (NLO) computations.

A class of processes where this program has been successfully completed
recently to NLO for a large number of observables is 
heavy quark
production\cite{giov},  
where the mass of the quark provides the hard scale which justifies
the applicability of perturbation theory. A comparison with experiment
is possible both for photoproduction (at HERA) and hadroproduction
(at the Tevatron), and quantities computed at NLO include total cross
sections, single-inclusive $p_T$ distributions, and $Q\bar Q$ 
correlations. A general feature of these results is a
considerable improvement of the NLO description of the data when
compared to the leading order one, even though there still remain
several instances in which the agreement is marred by large
uncertainties. Typical examples are  the normalization of
the total $c\bar c$ photoproduction cross section or of the
single-inclusive
$P_T$ distribution for $b$ production at the Tevatron (which  only
agrees with the theory for extreme choices of the scale $\mu$).

Progress in these computations has been achieved not only because of
the determination of the hard matrix elements for a large number of
processes, but also thanks to the development of efficient techniques
in order to go from the amplitudes to the physical cross sections
which correspond to physical observables. The main technical obstacle
here is that the infrared singularities due to the emission of
massless particles which plague the individual amplitudes only cancel
when combined into a physical  cross section. Whereas
in practice the cross section must be computed numerically in order to
compare to experiment, this
cancellation has to be carried out analytically before the numerical
calculation is feasible, by either excluding the singular regions from
the numerical integrations (slicing method) or subtracting the
analytically computed singularity before numerical integration in
order to make the latter finite (subtraction method). The main
disadvantage of the latter method, namely that it becomes cumbersome
in processes with many partons because of the large number of
individual cancellations, has been recently overcome thanks to a
systematic organization of the cancellations at the level of the cross
section~\cite{mastro}.

We are thus  reaching a stage where further progress requires new
theoretical ideas.
A representative selection of the problems we face, and the sort
of solutions we are after, is
the following: (a) we need more effective techniques to avoid the
proliferation of Feynman diagrams and terms in each diagram when
calculating amplitudes with many partons and loops; (b) we have to
cope with the divergent nature of the perturbative expansion in
$\alpha$ and consequently go beyond purely logarithmic accuracy; (c)
we must deal with processes with more than one hard scale; (d) we
must resum  logs, such as infrared logs,
even when  the   renormalization group doesn't help; (e) we would like
to extend the applicability of leading log resummation to cases, such
as logs of $1/x$, where the light-cone expansion is not available.

In the sequel we will give a brief overview of some of the 
techniques recently proposed 
to deal with all these issues.

\section*{Perturbation theory beyond Feynman diagrams}

The
determination of matrix elements with many partons is an important
ingredient in the computation of physically relevant QCD processes:
for example, the cross section for the production of six jets or five
jets and a vector boson in $p\bar p$ collisions is a significant background to 
top production. However the number of Feynman diagrams, and, more
importantly, the number of terms corresponding to each diagram grow
very rapidly with the number of partons involved. 

For example,
because the three-gluon vertex in QCD is the sum of six terms,
a typical  four gluon one-loop diagram (such as that where all gluons
attach to a gluon loop) will have about $6^4$ terms. Each of these
corresponds to a loop integral which, once evaluated, results by
itself in a lengthy expression, so the full starting amplitude
consists of about $10^4$ terms. Yet when the full computation is
carried through, the resulting amplitude~\cite{elsex} fits in less than
a page and is remarkably simple in structure. This large amount of
cancellation, together with the observation that each diagram does not
respect the symmetries of the final result, suggests that
Feynman diagrams are not an effective way of organizing the
computation.  

This also suggests that string theory might instead
provide a better way of organizing such calculations\cite{kosower}. 
Field theory can
be obtained from string theory by taking the limit of infinite string
tension, in which case all excited modes of the string decouple so
that only a field theory survives; by suitably choosing the string
theory one can make sure that the desired field theory is obtained in
the limit. The main reason why this
actually  simplifies the
calculation is that in string theory
at each order of perturbation theory there is only one diagram which
encompasses the full set of Feynman diagrams of the underlying field
theory: many cancellations thus take place
at an early stage of the computation. On top of this, string
amplitudes have several additional useful properties, such as, for
instance,  the fact
that it is possible to switch from bosons to fermions in the loop by
changing world-sheet boundary conditions, which allows using
information from one calculation to perform the other.

A significant obstacle which hampered progress in this direction was
the need to use a fully consistent string theory 
with the correct low-energy limit, such as the heterotic string. 
However, once computational rules have been derived,
the string theory can be  dispensed with. In fact,
it can be checked explicitly that the same rules can be
obtained from simpler albeit inconsistent theories such as the bosonic
string. One is then left with a set of rules which
can be used directly to compute generic one-loop helicity amplitudes.
These rules optimize and extend to the one-loop case a set of
ideas which had been developed (without resorting to string
techniques) to cope with the complexity of tree
level amplitudes with many partons, 
such as  spinor helicity methods to treat gluon
polarizations, color  decompositions, supersymmetric identities
and recursion relations between amplitudes with different numbers of
partons \cite{manpar}.

The rules can be further improved by use of unitarity methods. The
basic idea here is that the absorptive part of the one-loop amplitudes
can be determined by dispersion relations from the dispersive part,
which in turn is related to a product of tree amplitudes. The loop
amplitudes are thus computed by sewing tree amplitudes, rather than
tree diagrams, so that one can take advantage of the full set of
cancellations which have already occurred at the level of the
computation of the tree diagrams themselves.

While tree-level computations are now a closed chapter\cite{manpar}, 
since  
it is possible to determine numerically
and even (for certain helicities) analytically the amplitudes for any
number of particles, progress is made in one-loop
computations. Specifically, while the one-loop four-parton cross
sections were calculated using standard techniques~\cite{elsex}, the
corresponding helicity amplitudes have been first determined
exploiting string and unitarity methods, which have subsequently 
allowed the
determination of five-gluon amplitudes as well. The computation of
amplitudes with four partons and a vector boson is in
progress~\cite{kosower}. Also, with these techniques it 
has been possible to determine the
one-loop $2\to N$ gluon amplitude at the level of
next-to-leading logs of the center-of-mass energy of the gluon
pair \cite{vitt}.

Going beyond one loop is an interesting challenge, especially when one
realizes that no results beyond NLO are available for any process
depending on more than one kinematic variable: for instance, anomalous
dimensions for some specific operators are known beyond NLO, but no
splitting function is known beyond NLO. However, going beyond one loop
requires going back to the string theory as a guidance to formulate
suitable computational rules \cite{zvi,lor}. At present some results
in N=4 supersymmetric QCD are available~\cite{zvi}, while the
two-point Green function, from which the QCD $\beta$ function may be
extracted, should be available shortly~\cite{lor}. Perhaps more
interestingly, going to two loops suggests conjectures on the
structure of amplitudes which may lead to a determination of some
quantities (such as the $\beta$ function itself) to all
perturbative orders~\cite{zvi,lor}.

\section*{High perturbative orders beyond log accuracy}

It is a well-known fact that the 
perturbative expansion of gauge theories diverges. As a consequence,
perturbative computations cannot be pushed to indefinitely higher
orders. Nevertheless, one may  attempt a resummation 
of the divergent expansion. 
Even though a full understanding of the diagrammatic structure of QCD
is not available, individual towers of
diagrams which lead to  divergent behavior of perturbation
theory may be isolated. Specifically, the so-called renormalon diagrams, which
correspond to a vertex correction dressed by a chain of bubble quark 
self-energy insertions, lead to factorial behavior of the
corresponding perturbative coefficients\cite{beneke}. 

A factorially divergent series can be turned into  a geometric
 series by   Borel
 transformation:
if $C(t)$ admits an expansion
of the form $C=\sum_{n=1}^\infty n! t^{-n}$ then its Borel transform $B(u)$,
defined by the relation $C(t)=\int_0^\infty e^{-t u} B(u) du$,
satisfies  $B=\sum_{n=0}^\infty  u^n={1\over 1-u}$. If we wish to use this as a
 means to resum the original series we must invert the transform. We 
will then encounter
 a singularity on the path of integration at $u=u_0=1$ (infrared
 renormalon pole). If the series
 had alternating signs the singularity (ultraviolet renormalon) 
would not be on the path of
 integration, but the integral would still run outside the radius of
 convergence of the series; we will not discuss this case further.
The prescription chosen to
 treat the singularity introduces then an ambiguity of order
 $e^{-u_0t}=e^{-t}$. In typical QCD computations 
the expansion is in powers of
 ${1\over t}={1\over \ln(Q^2/\Lambda^2)}= \beta_0{\alpha_s\over 4\pi }$
 (to leading order) so the ambiguity is of order ${\Lambda^2\over
 Q^2}$.
This means that even though the expansion is in logs of $Q^2$, 
the resummation ends up
 producing terms which behave like powers of $Q^2$, so 
that a treatment of the
high-order behavior of the leading-twist perturbation theory requires 
going beyond leading twist.

The physics behind this phenomenon has  been extensively
elucidated\cite{beneke}:   the
chain of renormalon diagrams is sensitive to the infrared region of the
 loop momentum integration, which must be matched to the
power ultraviolet divergent behavior of higher twist operators which
appear in the operator-product expansion. Hence, this 
is a manifestation of operator mixing upon
shifts of the factorization scale beyond log accuracy.
The relevance of the infrared momentum integration region is
highlighted by realizing that
the sum of the same class of diagrams
can be obtained equivalently by simply computing a single vertex
correction, without the chain of bubble insertions, but with an
effective gluon propagator that corresponds to a massive
gluon\cite{beneke};  this
can also be viewed as the result of introducing a dispersive
representation for the strong coupling in terms of an effective
coupling which is regular in the infrared\cite{pino}. The poles in the
Borel transform discussed above are then in one-to-one correspondence with 
non-analytic contributions to the amplitude calculated with the effective
propagator. This equivalence is not exact, however, when calculating
renormalon corrections to less inclusive processes, in which the chain
of bubbles can directly contribute to the final state.

The relation between  power-like ambiguities of the perturbative
corrections to leading twist operators and their  
mixing with higher-twist contributions to the OPE
suggests that one may actually extract physical information from
renormalons. Indeed, any ambiguity related to operator mixing,
i.e. ultimately to the choice of factorization, must cancel in physical
observables. Thus, the renormalon ambiguity must be matched by an equal
and opposite ambiguity in the matrix elements of higher twist
operators. We may then conjecture that the $x$ dependence of the
renormalon ambiguity is the same as  that
of the higher twist contribution which corresponds to the 
set of these operators. Then,  by computing the renormalon
diagrams we can predict the shape of the higher twist contribution
up to a normalization.  This is
equivalent to
assuming that the matrix elements of
the higher twist operators are dominated by their ultraviolet
behavior (while
the particular subclass of
diagrams which corresponds to renormalons does indeed provide the dominant
perturbative behavior).

 This assumption doesn't have a solid theoretical foundation:  in fact,
it can only be approximately
true, because the logarithmic scaling violations of the higher
twist operators are not the same as those of the
renormalons\cite{akh}. However,  it seems to be
phenomenologically rather successful: the shape of the higher twist
contributions to structure functions obtained from a fit to the
observed scaling violations, for instance, display a good qualitative
agreement with those computed with this method\cite{pino,lech}. 
This suggests that indeed
the higher twist matrix elements are
dominated by their perturbative tail, and thus do not provide any
nonperturbative information on hadron structure\cite{beneke}. 

On the other hand, 
it also suggests that we may use the same method to predict the
shape of higher twist corrections even in cases where an OPE is not
directly available, such as  event shapes or jet 
observables\cite{beneke,pino,akh}. 
Performing the computation one arrives at  the interesting
result that such
processes may acquire $1\over Q$ corrections; this has been recently
shown to be the case (in a somewhat more subtle way) for fragmentation
functions as well\cite{benlor}.
Going one step further, one may assume that such $1\over Q$
corrections are universal\cite{akh,pino}. This assumption can again
be only approximately true; it may be justified in the
``dispersive'' approach (which, as already mentioned, is no
longer exactly equivalent to the Borel transform approach).

Recent progress in this field has involved the computation and
classification of such corrections for a large variety of
processes. The corresponding phenomenology (and the aforementioned
assumptions of perturbative dominance and universality) seem to work
much better than they ought to: this poses an interesting theoretical
problem. 
More unconventional recent research directions have involved the
study of nonlocal operators as a means to systematically identify the
kinematic regions related to the power corrections~\cite{sterren}, and the
use of lattice methods to compute numerically perturbative corrections
to the vacuum expectation value of the gluon condensate
to high orders and thus test whether they display the high order
behavior expected on the basis  of renormalon calculations, thereby
verifying directly  the dominance of renormalon diagrams on the full
perturbative behavior\cite{pinlat}. Interestingly, the observed large order
behavior  shows good agreement with the leading renormalon estimate, but only
after the lattice running strong coupling is redefined in order to eliminate a
$1\over Q^2$ term which should not be present according to
the OPE. The result suggests that
``spurious'' power corrections may arise if the strong coupling is not
properly defined beyond logarithmic accuracy\cite{rencoup}: this may provide a
window to the behavior of the running coupling beyond perturbation theory.

\section*{Heavy quarks beyond fixed order}

Resummation to all orders in the coupling of leading and subleading
logarithms  is routinely performed when solving the QCD evolution
equations for structure functions. This allows the resummation
to all orders of the logs
of the only large scale available when quark masses are neglected,
namely $Q^2$. It is clear however, that heavy quark
masses cannot be considered to be negligible in many physically
relevant applications: for instance, a large share of the  HERA $F_2$ data
comes from regions where $Q^2$ is close to the charm or the bottom
threshold.  

A standard way of including the heavy quark contributions is 
 to neglect them below threshold, while considering the heavy quark as
effectively massless above threshold (variable flavor number or VFN
scheme). However, 
above but not very far from the 
threshold this approximation is not  justified, and the threshold
behavior will be poorly reproduced by it. An alternative
possibility is to include the heavy quarks contributions 
(via photon-gluon fusion) to the hard coefficient functions,  while 
not including the heavy quarks among
the  active partons (fixed flavor number or FFN scheme). 
This, however, fails when $Q^2\gg m_Q^2$ because then
logarithmic terms in $\ln
{Q^2\over m_Q^2}$ are not resummed. 

The recent more precise  structure
function data, as well as the availability of direct determinations of
the charm parton distribution, 
call for an improved treatment. 
This can be achieved by means of $m\not=0$ factorization
theorems\cite{tung} (alternative approaches have also been
proposed\cite{mrrs}): 
the basic idea is that one may include in both the photon-gluon
fusion contribution to 
the coefficient function as well as the direct contribution from
the evolution of heavy quark partons above threshold, provided a
subtraction term is included to avoid double counting. 
The result then reduces to the
VFN scheme  when $Q^2>>m_Q^2$, and to the FFN scheme
when $Q^2\approx m_Q^2$, so that all relevant large logs are
being summed. The shortcoming of the method is its complexity:
 thus, in practice,
the most efficient scheme may depend on the specific observable. The
relevant isues, however, are settled from a conceptual point of
view, and  results  can now be used in phenomenological applications.

\section*{Resummation beyond the renormalization group}

QCD processes are typically marred by large infrared logarithms. The
standard renormalization group methods used to sum ultraviolet
logarithms are not applicable, yet resummation is required in order to
obtain accurate predictions.
For a
wide class of processes involving the emission of soft gluons this
resummation has been carried through. 
Consider, for instance, 
the top production cross section, whose accurate determination has
become recently of great phenomenological relevance. In
general,  emission of soft gluons has a small effect, because it
affects the kinematics of the process only slightly. However, 
when approaching the 
threshold for production of the  $\bar t t$ pair emission
of a soft gluon may have a large effect, because even the small amount
of energy spent in the gluon radiation can subtract a significant
fraction of the available energy, thereby suppressing the emission. 
Because  radiation effects are already included
in the  structure functions $f_i$
in eq.~(\ref{efact}) (when solving the QCD evolution equations)
the sign of the correction to the partonic cross section $\hat\sigma$
due to soft radiation
is scheme dependent:  in the $\overline{\rm MS}$ and DIS schemes $\hat\sigma$
the suppression is negative, i.e. $\hat \sigma $ is actually increased.

The emission of soft gluons is
thus 
logarithmically  enhanced: each extra emission produces a contribution
of order
$\alpha_s\ln^2(1-{Q^2\over s})$, where  
$Q^2$ is the invariant mass of the $\bar
t t $ pair ($Q^2=4 m_t^2$ if the pair is produced at rest). 
Close enough to threshold the log
 may offset the suppression due to $\alpha_s$, and
the soft gluon emission must be resummed to all orders in order to
achieve an accurate description of the process. 
This sort of
resummation has been achieved since some time for inclusive
processes, such as Drell-Yan or DIS, where it appears as the need to
resum to all orders logs of $1-x$ in the evolution equation, and it
has recently been accomplished to leading log accuracy
in the case of heavy quark production\cite{berg,cmnt,lsv}.

The leading log resummation is performed by  exponentiation in moment
space. Even though existing computations differ in the treatment of
subleading corrections, the results agree within error. The correction
is found to be rather small (the enhancement is smaller than $10\%$
and compatible with zero within the given uncertainties for all
calculations) but grows rapidly with energy due to the increase in
importance of the threshold region. The next-to-leading log corrections
have also been computed in moment space for the $\bar q q\to\bar Q Q$
and $gg\to\bar Q Q$ subprocesses\cite{sterkid}. 
The corresponding moment inversions
as well as the computation of the $gg\to gg$
amplitude are in progress\cite{sterkid}. 

In this context, phenomenology is lagging
behind theory: the relevant issues are well-understood
theoretically, and detailed analytic results are available, yet no systematic
phenomenology is available. Even at the fully inclusive
level, despite the increase of interest in accurate computations in
the large $x$ region related to the HERA events, the 
phenomenological relevance of resummation of $\ln(1-x)$ 
corrections to structure functions has not been systematically
included in available structure function analyses, and, for instance,
its impact on the determination of $\alpha_s$ from DIS is unknown.

\section*{The renormalization group beyond large $Q^2$}

A different class of large logarithmic corrections to perturbative
computations which calls for a resummation is that related to large
center-of-mass energy logs. In DIS these correspond to logarithmic
corrections  in
$1\over x$, which have of course attracted considerable interest since
the availability of DIS data at small $x$ from HERA.
The physics behind large energy logs is similar to the familiar physics
of the large logs of $Q^2$ which are summed by the
renormalization group: as the center-of-mass energy or, respectively,
the virtuality, increases so
does the phase space for parton radiation, and thus the latter is
enhanced by logs of the relevant scale. Specifically, in DIS
the value of $x$ in a given event
expresses the  longitudinal momentum which is
put on shell by the virtual photon, expressed in units of the incoming
proton's momentum. If the center-of-mass
energy of the $\gamma^* p$ collision $W^2=Q^2{1-x\over x}\approx
{Q^2\over x}$ is very large,
then  a very small fraction of it is put on shell by the virtual
photon. This implies that there is a large phase space for emission of a
parton cascade in which each parton's longitudinal momentum is very
soft as compared to its parent parton, so that the final (struck)
parton carries only a tiny fraction of the original proton's
longitudinal momentum. Each parton emission has a  logarithmic behavior at
the infrared edge of the longitudinal momentum integration ---
similar to that
discussed in the previous section, but now the soft
parton is that
which takes part in the cascade, i.e. the one which further fragments
until being eventually struck by the virtual photon, rather than that
which is emitted in the final state. As a consequence, each new
emission carries a factor of $\alpha_s\ln {1\over x}$.

The  leading energy log corrections to gluon-gluon
scattering have been determined  long ago\cite{bfkl,vittrev}, and recent
progress has involved making this consistent with the QCD evolution
equations by means of suitable factorization theorems,
as well as extending the computation to the quark sector\cite{ktfact}.
In this framework, leading logs of $1\over x$ appear as contributions 
proportional to ${1\over x}(\ln 1/x)^{k-1}$ to the $k$-loop
Altarelli-Parisi gluon splitting functions and the $k+1$--loop quark
splitting functions and coefficient
functions, whose coefficients are thus exactly determined  to all orders in
$\alpha_s$. Very recently, significant progress has been made towards the
determination of the next to leading corrections in the gluon sector,
in that the corrections to the anomalous dimensions have been
computed~\cite{bfkl}. Interestingly, part of the calculation has
been cross-checked by using the string-- and unitarity--based methods
which we discussed previously\cite{vitt}. 
In order for these result to be useful,
however, the corresponding coefficient function must still be
determined.

The theoretical and phenomenological status of summing large energy
logs by including $\ln {1\over x}$ contributions to the splitting functions
to all orders in the coupling remains however unsatisfactory. On the
one hand, the ensuing treatment of scaling violation is unnaturally
asymmetric, since leading logs of $1\over x$ are summed by including
them by hand in the anomalous dimensions, but the evolution equations
themselves resum logs of $Q^2$.
On the other hand, the inclusion of these terms in the usual evolution
equations is phenomenologically useless. While  the agreement between
the observed and computed scaling violations is spectacular at the NLO
level, it gets actually worse when energy logs are included 
unless one fine-tunes the  factorization scheme
in such a way that the energy logs have no detectable  effect\cite{roma}.

It has been suggested recently that both of these problems may be
overcome if the renormalization group is directly used to resum energy
logs\cite{noi}. The basic idea here is to construct a factorization
theorem (``energy factorization'')
in which, in comparison to the standard mass factorization,
the roles of  $Q^2$ and ${1\over x}$, i.e. of transverse and
longitudinal momentum scales, are interchanged. It is then
possible to obtain the resummation of all leading logs of $1\over x$
as the result of renormalization group improvement of the
energy-factorized cross section, 
by solving a leading-order evolution equation  with a splitting
function which depends on transverse momentum, and that can be easily
determined at leading order by means of standard Weizs\"acker-Williams
methods. While the coefficients of the leading $\alpha_s\ln {1\over
x}$
contributions to the DIS cross section 
are then the same as in the standard approach\cite{bfkl,ktfact},
energy factorization allows one to define   suitably energy--factorized
parton distributions, and to determine the leading order running of
the coupling. It is then found that there exists only one
energy-factorized parton, which at  large energy is asymptotically
free. This leads to a universal, perturbatively
calculable high-energy behavior of the inelastic scattering cross
section, which turns out to be consistent with unitarity bounds. The
most striking consequence of this approach is that it seems to imply
asymptotic freedom even when $Q^2$ is small provided the center of
mass energy is large enough, and thus  an unexpected
extension  of the kinematic domain in which  perturbative methods can be
applied.

\section*{Beyond perturbative QCD}

A common thread linking the directions of research which we have
sketched is that they lead to a  widening of the domain of applicability of
perturbative methods.
The availability of a perturbative approach to problems
(such as power corrections or small $x$ effects)  
which until recently could be exclusively tackled  by means of
effective models allows a cleaner separation of 
the perturbative physics from the
genuinely nonperturbative soft input. On the other hand,  the
extension of perturbative results to high orders and their  matching
to the underlying nonperturbative behavior may provide some clues on the
structure of the theory beyond perturbation theory. Thus the
availability of powerful and deep perturbative results leads us
naturally to look  beyond perturbation
theory itself\cite{genya}.

\bigskip
{\bf Acknowledgements:} I thank all the contributors to the working
group for their lively participation and for 
providing the ideas on which this paper is based,
E.~Levin for a fruitful collaboration in organizing the session,
and G.~Ridolfi and R.~D.~Ball for a critical reading of the manuscript.

\end{document}